\documentclass{article}

\usepackage{arxiv}

\usepackage[utf8]{inputenc} 
\usepackage[T1]{fontenc}    
\usepackage{hyperref}       
\usepackage{url}            
\usepackage{booktabs}       
\usepackage{amsfonts}       
\usepackage{nicefrac}       
\usepackage{microtype}      
\usepackage{lipsum}		
\usepackage{graphicx}
\usepackage{doi}
\usepackage{textcomp}
\usepackage{times}
\usepackage{setspace}
\usepackage{makecell}
\usepackage[font=small,compatibility=false]{caption}
\usepackage{listings}

\usepackage{amsmath}
\usepackage[mathlines]{lineno}
\usepackage[labelfont=bf]{caption}
\usepackage{float}
\usepackage[T1]{fontenc} 
\usepackage[utf8]{inputenc}
\usepackage{textcomp}
\usepackage{mathpazo}%
\usepackage{times}
\usepackage{setspace}

\title{Declining hydrologic function of coastal wetlands in response to saltwater intrusion}

\author{ \href{https://orcid.org/0000-0001-6382-1381}{\includegraphics[scale=0.06]{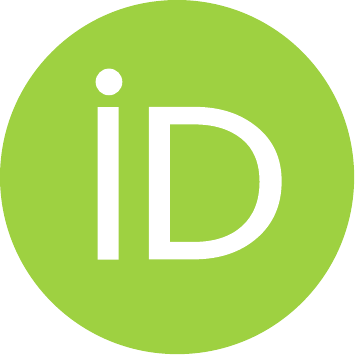}\hspace{1mm}Saverio Perri} \\
	High Meadows Environmental Institute,\\
	Princeton University, Princeton, NJ, USA.\\
		\texttt{sperri@princeton.edu} \\
	\And
	\href{https://orcid.org/0000-0003-3815-3929}{\includegraphics[scale=0.06]{orcid.pdf}\hspace{1mm}Annalisa Molini} \\
	Department of River-Coastal Science \& Engineering,\\
	Tulane University, New Orleans, LA, USA.\\
	\texttt{amolini@tulane.edu} \\
}

\renewcommand{\shorttitle}{\textit{arXiv}}

\hypersetup{
pdftitle={Declining hydrologic function of coastal wetlands in response to saltwater intrusion},
pdfsubject={q-bio.NC, q-bio.QM},
pdfauthor={Saverio Perri, Annalisa Molini},
pdfkeywords={Tidal wetlands, Coastal salinization, Plant salt tolerance, Ecosystem hydrologic function, Sea-level rise, Mangroves},
}

\begin{document}
\maketitle
\begin{abstract}
Salinity and submergence have shaped coastal wetlands into one of the most productive and yet fragile ecotones worldwide.
Sea-level rise alters both salinity and submersion regimes, threatening the future existence of these habitats.
Still, the magnitude of the impacts is elusive.
A current paradigm is that coastal communities will keep pace with sea-level rise through soil accretion, a process largely sustained by the interaction of plants and geomorphology.
This thesis, however, rests on the assumption that submersion is the main driver of bio-geomorphic processes and does not include the effects of salinity on plant hydrologic function.
Here, we show that salinity has a major influence on how tidal vegetation controls the depth of the water table --  a key ecohydrological process regulating plant establishment, survival rate, and productivity.
Our findings rely on long-term observations from the Florida Everglades and modeling results.
They indicate that altered salinity regimes can suppress transpiration, undermining in this way the ability of tidal ecosystems to control water table movements and contrast other forms of plant stress like waterlogging.
This effect is apparent in the mangroves of the Everglades' tidal fringe, where salinization has already yielded extensive decoupling between plant hydraulics and groundwater dynamics.
This mechanism leads to shallow water tables and poor soil-aeration conditions, with sizable impacts on coastal wetlands' hydrologic function and elasticity.
\end{abstract}

\keywords{Tidal wetlands $|$ Coastal salinization $|$ Plant salt tolerance $|$ Ecosystem hydrologic function $|$ Sea-level rise $|$ Mangroves}
\newpage

\section*{Introduction}
Tidal wetlands are a vital component of the global estuarine-coastal system~\cite{RamsarConventiononWetlands:2018,Bianchi2009}. 
They provide the outermost protection from rising sea levels, flooding, storms, and sea surge in coastal regions and low lands~\cite{costanza2021,Sun2020}. 
They control sedimentary dynamics~\cite{Feagin2009}, enhance water quality~\cite{Cheng2020,Verhoeven2006}, and support biodiversity~\cite{Denny1994}.
Here, plant communities have evolved to cope with a broad spectrum of environmental conditions, including shallow water tables~\cite{rodriguez2007} and large gradients of salinity and submersion~\cite{greb2006wetlands,saenger2002mangrove}.
The consequent emergence of a complex ecotone bridging salt-tolerant and freshwater communities has shaped coastal wetlands into one of the most productive and yet vulnerable ecosystems globally~\cite{Alongi2014,McLeod2011}, long-threatened by both natural and anthropogenic factors~\cite{Lovelock2015,Davidson2014,Pendleton2012}.
Only over the past century, about two-thirds of the world's coastal wetlands have been lost to human development, changes in land use, sediment starvation, and altered climatic and hydrological conditions~\cite{RamsarConventiononWetlands:2018,Davidson2014,Jankowski2017}. 
Despite recent efforts to contain human-induced wetland losses~\cite{Goldberg2020}, this trend could continue into the next decades, aggravated by accelerating sea-level rise~(SLR; \cite{Dangendorf2019}) and its most direct impacts on the ecological equilibrium of coastal wetlands: (a) increasing submersion, (b) alterations of the hydrological regime, and (c) salinization~\cite{Craft2009,Holmquist2021,Lovelock2015,Murray2022,Nicholls2004}.\\
\indent Still, coastal wetlands' response to SLR remains uncertain~\cite{Kirwan2016a,schuerch2018future,Wiberg2019}, with several authors pointing toward their further dramatic decline~\cite{Holmquist2021,Lovelock2015,Thorne2018,Craft2009} and many others even hypothesizing a future expansion~\cite{Kirwan2012,Rogers2019,schuerch2018future}.
In this last perspective, several studies have suggested that tidal wetlands could contrast rising sea levels through vertical~\cite{Kirwan2016a,Rogers2019} and lateral soil accretion~\cite{Kirwan2012,schuerch2018future} or expand through inland migration~\cite{Kirwan2016b,Kirwan2013}.
These results are based on the assumption that the feedback between plant productivity and geomorphology depends only on inundation depth and elevation~\cite{Kirwan2016a,Kirwan2010,morris2002,Mudd2009}, while other sources of plant stress, like salinity, are rendered as an implicit result of submersion.
However, as SLR enhances seawater intrusion and coastal salinization~\cite{Nordio2022}, the impacts of salinity on plant-water relations~\cite{bui2013soil,perri2017salinity,perri2018plant} cannot be neglected.
Failure to consider these effects -- which are not captured in current models -- may result in incorrect estimates of coastal wetlands' organic carbon accumulation~\cite{Craft2007,Neubauer2008}, a major component of soil accretion~\cite{Morris2016,Nyman2006,Turner2002}, and eventually overemphasize coastal wetlands' future resilience.\\
\indent Previous appraisals have focused on salinity's most direct and local effects on plant growth and productivity~(see Ref. \cite{Herbert2015}). 
The majority of these studies are ecosystem-specific and did not examine a key aspect of the problem -- the hydrological nature of salinity controls~\cite{Bathmann2021,Perri2020pnas,Perri2019}. 
In what follows, we show that salinisation does affect the hydrologic function of coastal wetlands by regulating the interaction between plant communities and the water table, with impacts already discernible in observations. 
We interpret this mechanism by introducing the limiting effects of salinity on plant transpiration and hydrological partitioning~\cite{perri2018plant,Perri2019}. 
We also demonstrate that the strength of these controls largely depends on plant salt tolerance~\cite{perri2018plant,Perri2020pnas,Perri2019}, with potential impacts on tidal ecosystems resilience, species competition, biogenic accretion processes, and the overall adaptability of coastal wetlands to rising sea levels.
\section*{Observed effects of salinity on plant-water table interactions}\label{sec:observations}
Salinity limits plant transpiration both through osmotic effects -- which lower soil water potential and impair water uptake from roots -- and ionic stress, which kicks in when salt within plant tissues reaches toxic levels~\cite{munns2008mechanisms}. 
To cope with the effects of salt stress, plants have developed a wide range of adaptations that define their salt tolerance. 
The main function of salt tolerance is to allow plants to contrast the effects of salt stress up to a species-specific salinity threshold.  
Above this threshold, transpiration rapidly declines, and plant hydrologic function is substantially impaired~\cite{munns2008mechanisms}.
At the same time, transpiration is how vegetation regulates the depth of the water table in both tidal~\cite{Dacey1984} and freshwater wetlands~\cite{Dube1995,Roy2000}.\\
\indent Here, we want to test the hypothesis that salinity controls are strong enough to affect the way vegetation governs groundwater dynamics.
We also want to verify whether these controls depend on salt tolerance and plant hydraulic design.
\begin{center}
\begin{figure}[H]
\includegraphics[width=37pc]{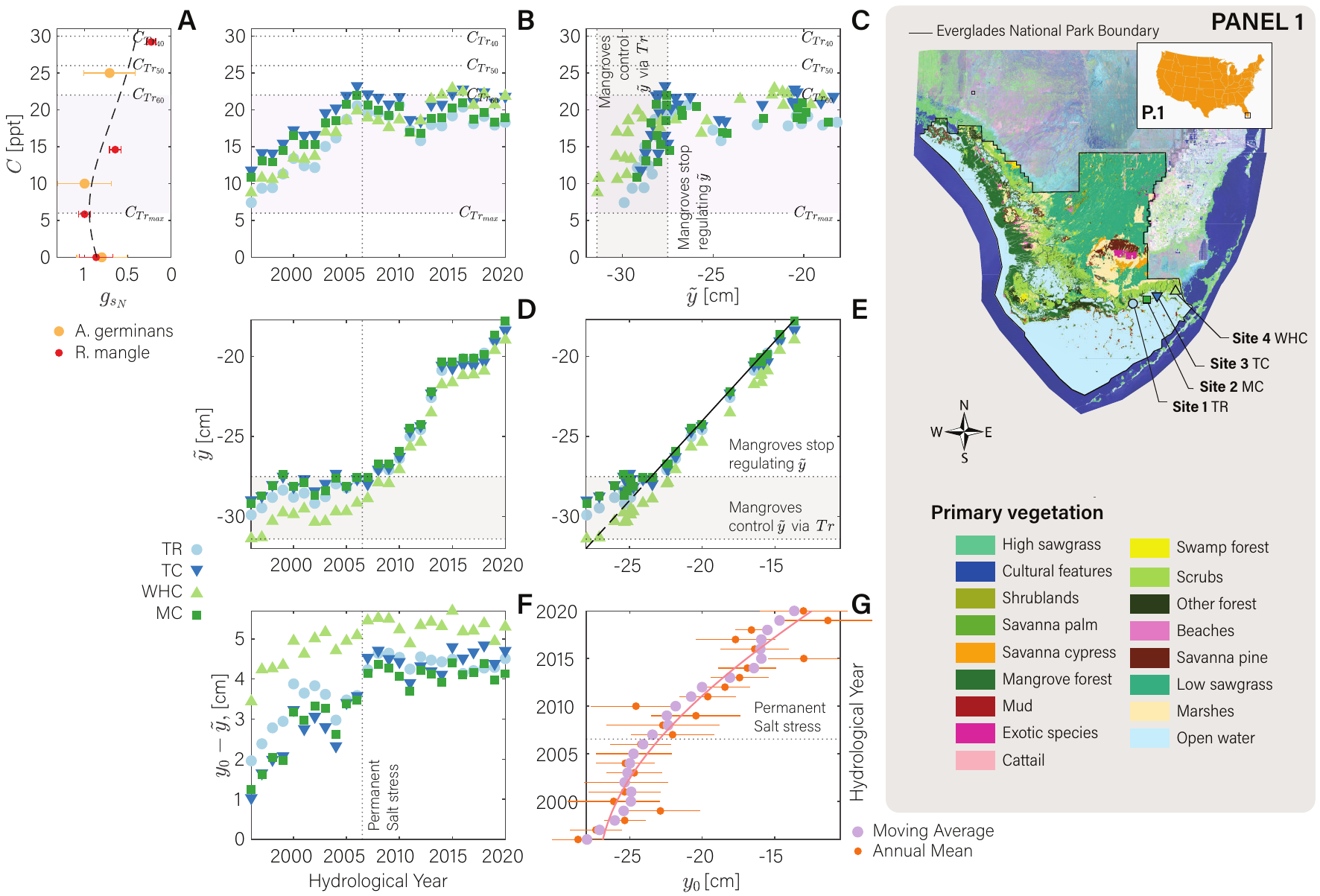}
\caption{\textbf{Observed feedback between salinity and water table depth.} \textbf{Panel 1}, Right side, Geographic location of the Florida's Everglades USGS monitoring stations used in the study. 
The background map is a composite of the Southern Florida LandSat 7 mosaic and Everglades' vegetation map from the Everglades Depth Estimation Network (EDEN) project.
\textbf{a}, Normalized stomatal conductance $g_{s_N}$ as a function of salinity $C$ for \textit{Rhizophora mangle}~\cite{Lin1993} (small red dots) and \textit{Avicennia germinans}\cite{Suarez2006} (large orange dots).
Error bars represent the standard deviation over multiple experimental trials at fixed salinity. The general relation between $g_{s_N}$ and $C$ is obtained as a non-linear fit across the two species (black dashed line, $R^2=0.85$).
\textbf{b},\textbf{d} and \textbf{g} temporal variability of salinity $C$ and water table depth $\tilde{y}$ for the four stations in \textbf{a}, and sea level $y_0$ at Key West, FL NOAA tidal station. 
To isolate long-term climatological trends, the temporal variability of $C$, $\tilde{y}$, $y_0$ (large purple dots in \textbf{g}) and $\Delta{y}$ (\textbf{f}) is represented as a centred four-year moving average of the annual means (see Methods).
Panel~\textbf{g} also shows the $y_0$ unfiltered annual mean (small orange dots) plus/minus its intra-annual standard error, overlaid to the best non-linear fit (sum of sine, $R^2=0.88$; thin red line). 
The $y_0$ does not account for subsidence or sediment compaction (absolute sea level).
For comparability, both $\tilde{y}$ and $y_0$ are referenced to the North American Vertical Datum 1988 (NAVD88; Methods).
\textbf{c}, Water table depth $\tilde{y}$ variability as a function of salinity $C$.
\textbf{e}, Relation between $\tilde{y}$ and $y_0$.
\textbf{f}, Temporal trend of the lateral gradient between the sea surface and the water table depth $\Delta{y}$ over the observation period.}
\label{fig:Fig_1}
\end{figure}
\end{center}
We consider the case of salt-tolerant plant communities populating the Southeastern fringe of the Everglades National Park.
These are scrub mangrove forests~\cite{Ewe2006,Poret2007} dominated by the species \textit{Rhizophora mangle} and \textit{Avicennia germinans} -- two facultative \textit{halophytes}~\cite{Wang2011} with similar salt tolerance characteristics (\textit{Materials and Methods}).
Our analysis is based on four micro-tidal sites~\cite{Davis2001} from the U.S. Geological Survey network (Taylor River at Mouth, Mud Creek, Trout Creek, and  West Highway Creek; \textit{PANEL 1} in Fig.~\ref{fig:Fig_1}) with similar hydrological attributes (seasonal precipitation and wind-driven estuarine input) but slightly varying vegetation and salinization characteristics.
We concentrate on belowground water tables -- accounting for more than 97\% of the available observations -- and exclude submersion periods during which flooding may represent the dominant source of plant stress.\\
\indent Continuous observations of soil pore water salinity and belowground water table depth are available at the four sites for the hydrological years 1996-2020 (May to June of the following year) with native sub-hourly resolution (\textit{Materials and Methods}).
The effects of salinity on mangrove transpiration are quantified through the relation between normalized stomatal conductance $g_{s_N}$ -- a proxy for plant transpiration -- and salinity $C$ (Fig.~\ref{fig:Fig_1}\textit{A}).
This relation is generalized through a non-linear fit of published experimental data collected under controlled conditions for both \textit{R. mangle}~\cite{Lin1993} and \textit{A. germinans}~(black dashed line \cite{Suarez2006}). 
Analogously to other salt-tolerant species~\cite{perri2018plant,munns2008mechanisms}, \textit{R. mangle} and \textit{A. germinans} show a non-monotonic relation between salinity and conductance. 
I.e., their transpiration rate is generally higher at an intermediate salinity (the salinity of maximum conductance, $C_{Tr_{\rm{max}}}$) than in freshwater conditions~($C_0$, \cite{munns2008mechanisms}).
Above $C_{Tr_{\rm{max}}}$, $g_{s_N}$ and transpiration decline -- with $C_{Tr_{60}}$, $C_{Tr_{50}}$ and $C_{Tr_{40}}$ (Fig.~\ref{fig:Fig_1}\textit{A}--\textit{C}) representing the salinity at which maximum transpiration is reduced by $40$\%, $50$\% and $60$\% respectively -- and plant communities gradually lose the ability to regulate water table depth and soil aeration.\\
\indent Fig.~\ref{fig:Fig_1}\textit{B},~\textit{D} and \textit{G} show multi-decadal trends of soil pore water salinity, $C$, belowground water table depth ($\tilde{y}$; see Methods) and sea level, $y_0$ -- a key forcing of lateral flows -- at the four sites.
While $y_0$ displays a continuous non-linear increasing trend over the study period, the behavior of $C$ and $\tilde{y}$ appears more complex.
Salinity first increases up to a saturation threshold $C \sim C_{Tr_{60}}$ to stabilize around that value afterward (2006 onward).
In contrast, $\tilde{y}$ remains approximately stationary while $C$ increases and rises rapidly hereafter.\\
\indent Recent studies indicate that increasing sea levels are one of the leading drivers of the shallowing of coastal water tables~\cite{Maliva2021} and related this mechanism to an increased risk of coastal flooding~\cite{Befus2020}.
Consistently with these studies, $\tilde{y}$ and $y_0$ at the four sites (Fig.~\ref{fig:Fig_1}\textit{F}) show a significant linear correlation (correlation coefficients varying between 0.95 at TC and 0.98 at WHC, significant at the $99$\% confidence level).
However, these strong correlations mainly arise from data for the period 2007 to present, whereas the relation among $\tilde{y}$ and $y_0$ is, in general, non-linear (Fig.~\ref{fig:Fig_1}\textit{E}).
Between 1996 and 2006, the shallowing of the water table with $y_0$ is negligible, while after the hydrologic year 2006, $\tilde{y}$ rises at a similar quasi-linear pace with sea level $y_0$ (Fig.~\ref{fig:Fig_1}\textit{D} and \textit{SI Appendix} Fig.~S1).\\
\indent This ``hokey-stick'' pattern can be explained by bringing to the picture the non-monotonic relation between $g_{s_N}$ and salinity in Fig.~\ref{fig:Fig_1}\textit{A}.
As sea levels rise, it is possible to identify two distinct phases in the dynamics of the water table.
During the first phase, $C<C_{Tr_{60}}$ (a low to moderate salt stress phase) and vegetation can maintain relatively deep and stable water tables by transpiration -- i.e., plant communities can control $\tilde{y}$ beyond the forcing of SLR (Fig.~\ref{fig:Fig_1}\textit{D}).
However, since sea level $y_0$ keeps increasing, vegetation controls on groundwater dynamics enhance the difference in elevation between the sea surface and the water table, $\Delta y=y_0-\tilde{y}$ which is also the main driver of seawater intrusion (Fig.~\ref{fig:Fig_1}\textit{F}).
This last mechanism accelerates coastal salinization (see Fig.~\ref{fig:Fig_1}\textit{B}) and, below the permanent salt-stress, represents for salt-tolerant species a tool to engineer the environment in their favor.
An analogous ecohydrological control has been observed in salt-affected drylands, where \textit{halophytes} enhance soil salinization to their advantage by sustaining higher transpiration rates -- and a lower frequency of leaching events -- than salt-sensitive species~\cite{Perri:2018ur,perri2022natgeo}.
However, with salinity approaching the threshold of salt stress ($C \sim C_{Tr_{60}}$), plant transpiration declines substantially, and vegetation controls on the water table weaken.\\
\indent The second phase commences when $C_{Tr_{60}}$ is exceeded, and salt stress becomes the predominant state.
Salt stress brings about the transition from a stationary ($\tilde{y}\simeq$ constant) to a non-stationary ($\tilde{y}\propto y_0$) groundwater dynamics.
Transpiration is substantially impaired, and the control of plants on the water table becomes gradually marginal.
Plant function and hydrology are now decoupled.
During this phase, the relation between $\tilde{y}$ and $y_0$ becomes approximately linear (Fig.~\ref{fig:Fig_1}\textit{F}).
As a result, both lateral flows and saltwater intrusion reach steady-state, preventing the further advancement of salinization (Fig.~\ref{fig:Fig_1}\textit{B} and \textit{F}).
The two phases are even more marked in Fig.~\ref{fig:Fig_1}\textit{C}, representing the relation between $C$ and $\tilde{y}$ ($\tilde{y}$ shown on the x-axis for visualization purposes).
Here, $\tilde{y}$  varies slowly till salinity remains within the limits of the optimal transpiration range and becomes rapidly shallower once $C_{Tr_{60}}$ is permanently exceeded.\\
\indent Salinity thus drives the transition to new forms of plant stress like hypoxia through a gradual loss of ecosystem resilience.
The resulting shift to a state of impaired plant hydrological function potentially paves the way to further ecosystem shifts -- like the transition to open water ($\tilde{y}\approx y_0$) -- with a pattern resembling other critical transitions captured in terrestrial and marine ecosystems~\cite{Scheffer2001}.  
\begin{center} 
\begin{figure}[H]
\includegraphics[width=38pc]{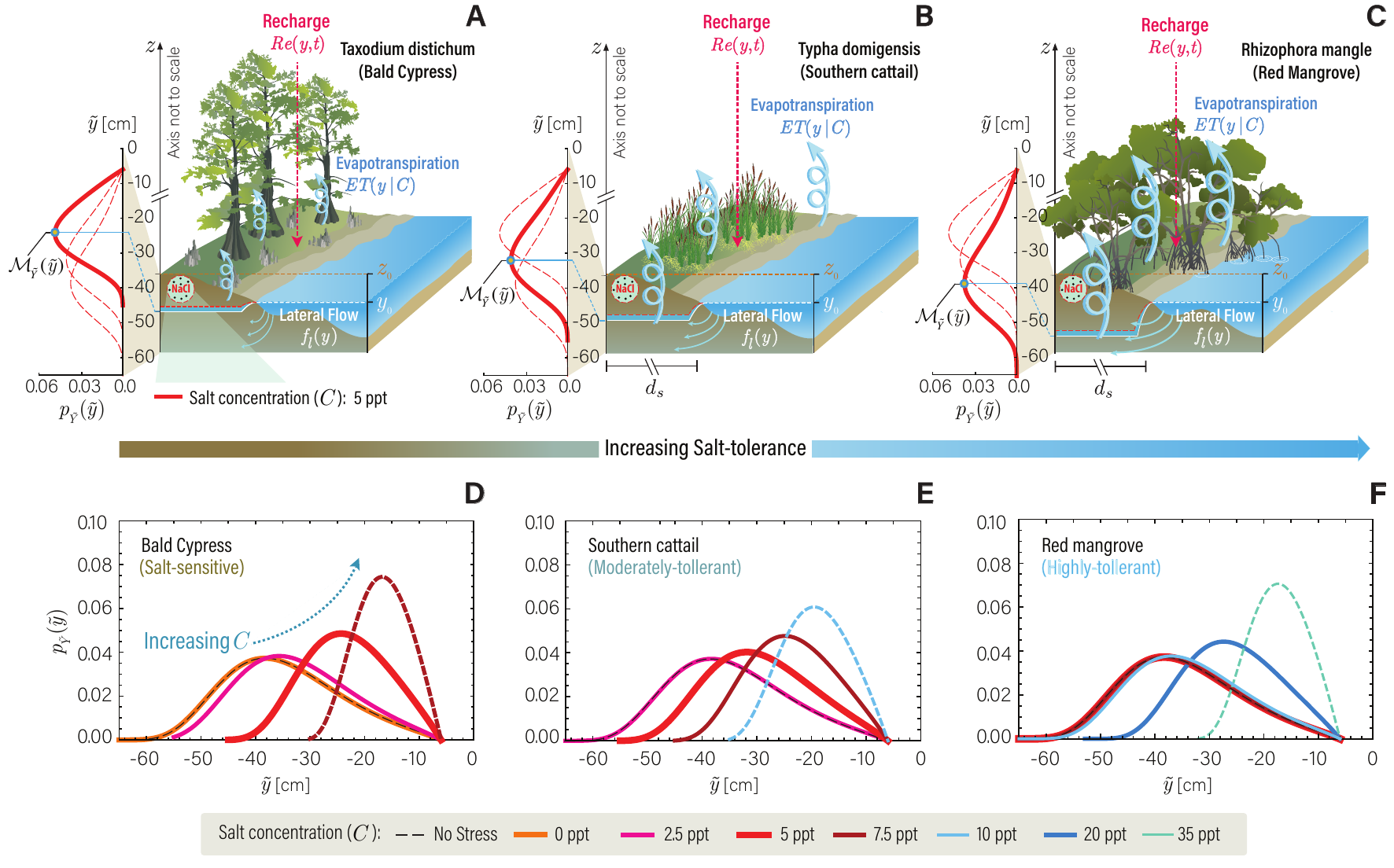}
\caption{\textbf{The role of salt tolerance.} (\textit{A}--\textit{C}) Conceptual representation of how salinity regulates water table depth depending on dominant species salt tolerance: (\textit{A}) salt-sensitive (Bald Cypress, \textit{Taxodium distichum}), (\textit{B}) moderately-tolerant (Southern cattail, \textit{Typha domigensis}), and (\textit{C}) highly-tolerant species (Red mangrove, \textit{Rhizophora mangle}). The probability density function (pdf) of the water table depth $\tilde{y}$ is obtained by solving the tidal soil water balance at steady state for soil parameters characteristic of the South-eastern coast of Florida's Everglades (\textit{Materials and Methods}) and conditional to a fixed water salinity $C$ of 5 ppt (solid red line). The median of $\tilde{y}$, $\mathcal{M}_{\tilde{Y}}(\tilde{y})$, characterizing each ecosystem is also shown in the diagrams (z-axis and distance from the shoreline, $d_s$, not in scale).
(\textit{D}--\textit{F}) Pdf's of $\tilde{y}$ for the three representative species in (\textit{A}--\textit{C}) and for different levels of salinization $C$, ranging between freshwater ($0$ ppt) and average sea-water ($35$ ppt) conditions. Bold dashed lines refer to the salinity for which each species is likely to suffer elevated mortality rates; black thin dashed lines denote the absence of salt stress for the considered species.
}
 \label{fig:Fig_2}
\end{figure} 
\end{center}
\section*{A simple interpretation of salinity ecohydrological role}
\label{sec:salinitymodelling}
To diagnose whether salinization is the main driver of the trends observed in Fig.~\ref{fig:Fig_1}, we introduce a salinity-dependent model of the soil water balance in the tidal fringe.
The model accounts for shallow water table conditions and random groundwater recharge (\textit{Materials and Methods}). 
Here, shallow water tables are assumed to represent coastal areas that -- similarly to our sites -- first experienced the effects of saline water intrusion.  
The model is used to explain the impacts of salt stress on the equilibrium dynamics of the water table depth, $\tilde{y}$, and expands previous formulations of the stochastic soil water budget in groundwater-dependent ecosystems where salinity effects were not considered~\cite{Pumo2010,Laio2009,Tamea2009,Tamea2010}.
Parameterizations are set to reproduce plant-water relations at the plot scale and across a homogeneous soil column.
The effects of salinity on plant-soil water relations are introduced through the dependence of evapotranspiration, $ET$, on pore water salinity, $C$, and the specific response of different species to salt stress -- i.e., plant salt tolerance.\\ 
\indent Following the approach proposed in previous studies~\cite{Perri:2018ur,Perri2020pnas,maas1999crop}, we render the relation between plant transpiration, $T_R$, and salinity as a step-wise linear relation, where $T_R$ is unaffected by salinity up to a species-specific concentration threshold, $C_T$ (a proxy of $C_{Tr_{max}}$ in Fig \ref{fig:Fig_1}), and declines above $C_T$ with slope $\phi \beta$. 
The parameter $\phi$ represents the ratio between $T_R$ and $ET$, while $\beta$ is the rate of decay of $T_R$ with salinity, also determined by species-specific physiological traits~(\cite{Perri:2018ur,Perri2020pnas}, and \textit{Materials and Methods}).
For a given level of salinization, the water table depth, $\tilde{y}$, is then governed by the balance between the salinity-limited evapotranspiration losses, $ET(y|C)$, the stochastic groundwater recharge, $Re(y,t)$, and lateral flows, $f_l(y)$. 
The parameters $\{\beta,C_T\}$ represent species-specific salt tolerance, and the different terms of the balance depend on the separation surface between the saturated and the unsaturated soil, $y(t)=\tilde{y}-\psi_s$, with $\psi_s$ the air entry tension~\cite{Laio2009}. 
Random inputs of precipitation determine the recharge, $Re(y,t)$.
\begin{center}
\begin{figure}[H]
\centering
\noindent\includegraphics[width=38pc]{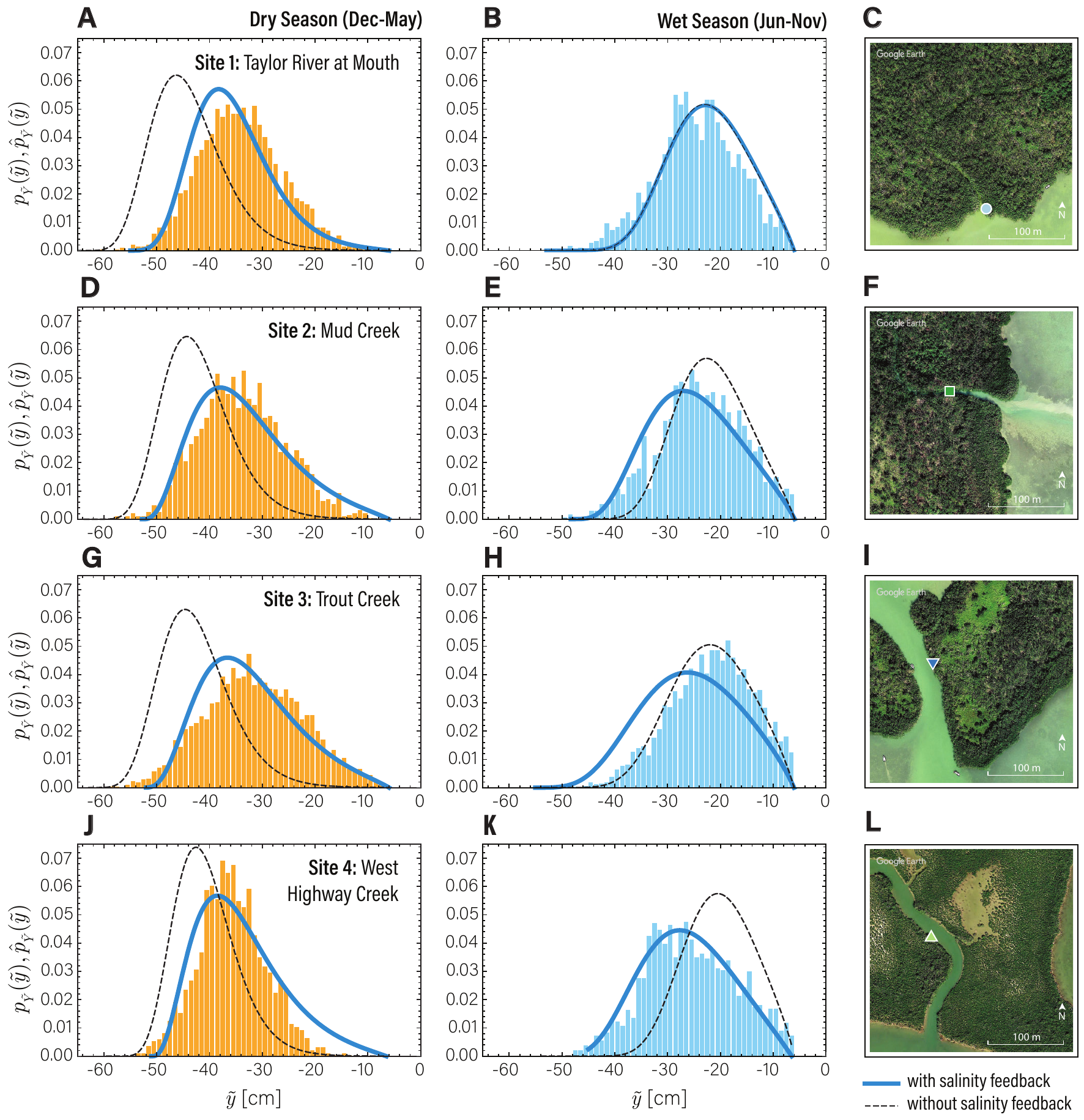} 
\caption{\textbf{Modeled effects of salinity on water table dynamics} Theoretical pdf's of the water table depth, $\tilde{y}$, for the dry (December through May) and the wet (June to November) season compared with the empirical distributions at Site 1 (Taylor River at Mouth; \textbf{a},\textbf{b}), Site 2 (Mud Creek; \textbf{d},\textbf{e}), Site 3 (Trout Creek; \textbf{g},\textbf{h}) and Site 4 (West Highway Creek; \textbf{j},\textbf{k}) over the period 1996-2006. Bold continuous lines corresponds to the salinity-dependent soil water balance, while dashed lines refer to the pdf's in absence of salinity effects. \textbf{c}, \textbf{f}, \textbf{i} and \textbf{l}, Forest structure zoom-in's at the four observation sites as obtained from Google Earth v. 7.3.3.7721 imagery.}
\label{fig:Fig_3}
\end{figure}
\end{center}
\indent This zero-order approach allows us to investigate how the introduction of salinity-limited evapotranspiration losses modifies the probability of attaining deep water tables in tidal wetlands.
Fig. \ref{fig:Fig_2}\textit{A}--\textit{C} shows a conceptual representation of the water balance of coastal wetlands for a relatively low salinization level ($C=$5 ppt) and species of increasing salt-tolerance: \textit{Taxodium distichum} (Bald Cypress; \textit{A}), \textit{Typha domingensis} (Southern cattail; \textit{B}), and \textit{R. mangle} (Red Mangrove; \textit{C}) -- commonly found along Everglades' salinity gradients ~\cite{richardson2008everglades}.
For the different species, we also show the modeled probability density functions (pdf), $p_{\tilde{Y}}(\tilde{y})$, of the water table depth conditional to values of salinity ranging between freshwater (0 ppt) and average seawater salinity (35 ppt; Fig.~\ref{fig:Fig_2}\textit{D}--\textit{F}).
The model can capture previously observed salinity controls, offering at the same time a new perspective on species competition along a salinity and salt tolerance gradient.\\
\indent If across the different species, increasing salinization always produces shallower water tables, \textit{halophytes} like the mangroves populating the tidal fringe can maintain a deeper water table than \textit{glycophytes} (salt-sensitive species). 
Through this mechanism, \textit{halophytes} further enhance seawater intrusion and soil salinization, re-engineering the environment in their favor~\cite{passioura1992mangroves,Wendelberger2017,Perri:2018ur}. 
This mutually reinforced feedback between halophytic species establishment and salinization can be seen as an additional hydrologic forcing on species succession, with apparent implications for the restoration of freshwater coastal wetlands. 
We next investigate whether the model can quantitatively reproduce the probabilistic features of the observed water table depth.
Fig. \ref{fig:Fig_3} shows the empirical distributions $\hat{p}_{\tilde{Y}}(\tilde{y})$ of the water table depth at the four study sites for the dry (December-May) and wet (June-November) seasons. We use here the observations relative to the hydrological years 1996-2006, for which $\tilde{y}$ is approximately stationary.
These are contrasted with the modeled pdf's in the absence of salinity effects (black dashed line), and for a salinization level, $C$ (bold continuous blue line) inferred from observations (\textit{SI Appendix}).
Examining the seasonal probabilistic structure of the belowground water table helps us further separate salt stress's effects from submersion, hypoxia, and hydrodynamic drivers.\\
\indent Consistently with previous studies~\cite{Pumo2010}, the model that does not account for salinization seems unable to reproduce the observed distribution of $\tilde{y}$.
However, if the limiting effects of salinity are introduced, the theoretical distributions well-capture the observed probabilistic structure of the water table.
The agreement is particularly marked for the dry cold season (Fig.~\ref{fig:Fig_3} \textit{A}, \textit{D}, \textit{G}, \textit{J}) when salinity represents the dominant source of abiotic stress for Mangroves, and salt stress is in anti-phase with submersion.
Here, salinity gradually suppresses transpiration and, with it, the capability of plants to regulate groundwater dynamics, resulting in water tables substantially shallower than the ones predicted by the stochastic water budget model in the absence of salt stress.
In contrast, the differences between the two models are less notable during the warm wet season, when salinity decreases due to major freshwater inputs, seawater thermal expansion is peaking, and the effects of submergence, local hydrodynamics, and soil anisotropy gain a dominant role.
Still, in the predominantly saturated conditions of coastal wetlands' soils, salinization appears to be the main source of water stress, able to control hydrologic function even across the most resilient halophytic plant communities like the Mangroves of the tidal fringe.
As salt exposure becomes more frequent and prolonged, tidal systems are likely to move from a salt stress-dominated phase (similar to the one observed in the years 1996-2006 in Fig.~\ref{fig:Fig_1}) to a submergence-dominated phase, further enhanced by the decoupling between vegetation and the water table.
\section*{Synthesis and implications}
\label{sec:discussion}
We have shown that salinity exerts a key hydrologic control on how plant communities interact with the water table.
Additionally, we demonstrated that species-specific salt tolerance could determine the strength of this interaction. 
Coastal salinization hence translates into a form of water stress that coastal wetlands are experiencing on top of other stressors like submersion, and which is still missing in current models.
We mainly concentrated on how sea level variability and hydroclimatic drivers determine seawater intrusion over large (climatological) temporal scales.\\
\indent In this context,  the pace of salinization is regulated by climate variability and plant function through $\frac{d \tilde{y}(t)}{d t}$ and by the sea level, through $\frac{d y_0(t)}{d t}$.
If the rate at which the water table rises is lower than the sea level, salinization is enhanced and vice versa.
A clear link between salt stress and water table dynamics emerged from both observations (Fig.~\ref{fig:Fig_1}) and the modeling exercise in Fig.~\ref{fig:Fig_2}--\ref{fig:Fig_3}.
Yet, factors other than SLR may have contributed to shaping the observed trends and accelerating salinity-driven critical transitions in our study area.\\
\indent Extreme events like hurricanes and prolonged droughts have been connected with intense episodic pulses of salinization~\cite{Lagomasino2021} and could have enhanced the loss of hydrological function of the Mangrove in the tidal fringe.
Three major hurricanes interested the Everglades during our study period: Katrina (August 2005), Wilma (October 2005), and Irma (September 2017).
Although their impacts were predominantly felt along the Southwestern coast of the Everglades, we can expect that salinization dynamics over the study region were partially impacted by saltwater incursion.
Similarly, `dry' hydrologic years with below-average precipitation like 2004 (July 2003 to June 2004; \textit{SI Appendix}, Fig.~S2) could have enhanced salinization and its effects on the vegetation-hydrology coupling.\\
\indent The rare-event component of the salinization dynamics is challenging to capture through a simple water budget approach.
However, it is interesting to note that the rapid submersion of the southern Florida coastline during the late Holocene has been related to the increased frequency of protracted droughts and extreme storms rather than the direct effects of SLR~\cite{jones2019}.
In light of the connection between tropical storms (and droughts) and salinity, this result could be interpreted as a further indication of the hydrological role of salinity in enhancing the effects of SLR and promoting ecological shifts in coastal wetlands.
Similar transitions, in turn, have irreversible impacts on coastal wetlands' elasticity and their potential to contrast SLR through vertical biogenic accretion.
If, as an example, prolonged droughts can already enhance soil pore water salinity when compounded with rising sea levels and sustained seawater intrusions, their effects are potentially amplified.
As a form of water stress, salinity has a crucial ecohydrological function in shaping coastal wetlands under SLR. 
This function should be explicitly considered in assessing the future resilience of coastal systems and in designing restoration measures.
\section*{Data and Methods}
The controls of salinity on the hydrologic function of coastal wetlands were explored using historical data of water table depth and pore water salt concentration combined with a stochastic model of water table dynamics. The model explicitly accounts for the limiting effect of salinity on plant water uptake and the average hydroclimatic forcing regulating the long-term dynamics of the water table depth in tidal wetlands. 
\subsection*{Sites description}
\label{sec:sites}
We used hydrologic observations from four U.S. Geological Survey (USGS) tidal sites located in the Southeastern Florida Everglades (Fig.~\ref{fig:Fig_1},~\textit{PANEL 1}). 
From West to East, they comprise the stations of Taylor River at Mouth (Site 1; USGS 251127080382100), Mud Creek (Site 2; USGS 251209080350100), Trout Creek (Site 3; USGS 251253080320100), and  West Highway Creek (Site 4; USGS 251433080265000).
The four sites are part of the Everglades Depths Estimation Network (EDEN).
They are characterized by similar flat topography (about 10 cm above MSL), hydrological regime, and proximity to the sea, but moderately different vegetation types (i.e., different percentages of the dominant mangrove species \textit{R. mangle} and \textit{A. germinans}), vegetation coverage and average soil water salinity. 
Southeastern Everglades' mangroves represent plant communities already massively exposed to coastal salinization but less impacted by sporadic disturbances like hurricanes than mangroves along the Western and Southwestern coastline~\cite{Lagomasino2021,Smith2009}.
The Taylor River (West) and the C-111 drainage canal (East) represent the main freshwater inputs to the study area ~\cite{langevin2005simulation,satbenau2012salinity,Marshall2014}. 
Here, precipitation and winds are the two major hydrologic drivers, while the influence of tides is limited~(micro-tidal regime; \cite{Davis2001}). 
While the Everglades have undergone nearly 150 years of water flow management through drainage, channelization, and flood control~\cite{national2019progress}, the regulation of freshwater inputs has predominantly affected less salt-tolerant inland communities~\cite{langevin2005simulation,national2019progress}. The four coastal sites considered in this study have been only marginally impacted by recent regulation efforts and, overall, in a comparable way.
\subsection*{Environmental and ecological drivers}
\label{sec:edrivers}
The Everglades are characterized by a subtropical climate~\cite{Peel2007}. About 80$\%$ of the annual precipitation occurs during the wet season from June to November, while the dry season extends from December to May~\cite{Duever1994}. 
The marked divide between wet and dry seasons translates into a strong signature in both the water table dynamics and salinity variability, with shallower water tables (and lower salinity) during the wet season and deeper water tables (and elevated salinity) throughout the dry season~\cite{Ewe2007,Pumo2010}.
In contrast, sea-level $y_0$ is maximum during the summer (and the wet season), as a consequence of local thermal expansion, and minimum in winter~(dry season; \cite{Wahl2014}).
At the selected sites, the water table is predominately below the ground level for most of the year, with only 7-9 days/year of water level above the soil surface (waterlogging).
The dominant vegetation at the Taylor River at Mouth site (Site 1) consists of the mangrove \textit{R. mangle} (Red mangrove) in its dwarf variant~\cite{Koch1997,Barr2009}. 
In contrast, the Mud Creek, Trout Creek, and West Highway Creek sites are characterized by a lower freshwater input and higher salinity, supporting more salt-tolerant species~\cite{Loveless1959,Ross2000}. 
Whether \textit{R. mangle} can still be found in the lower tidal zone at these sites, \textit{A. germinans} and shrubs species represent the dominant vegetation in the upper zone~\cite{Egler1952}.
\subsection*{Data}
\label{sec:salinitywtd}
At the four selected sites, we obtained water table depth ($\tilde{y}$; as referenced to the North American Vertical Datum of 1988, NAVD88), and pore-water salinity ($C$) data over the period 1996-2021. 
The native temporal resolution of the observations is 15 minutes.
Sub-hourly data were then aggregated at the daily scale and used to obtain the $\tilde{y}$ seasonal empirical distribution shown in Fig.~\ref{fig:Fig_3}. 
Annual statistics based on the hydrological year running from June to the following May were also computed. 
Monthly sea-level, $y_0$, observations for the years 1996--2021 were obtained from the National Oceanic and Atmospheric Administration (NOAA) at the Key West tidal station (KYWF1 -- 8724580). For comparability with $\tilde{y}$, $y_0$ is also referenced to the NAVD88 datum.
To infer the trends in water table depth, salinization, and sea-level rise, we computed the 4-years centered moving averages of the annual $y_0$, $\tilde{y}$, and $C$ average values (Fig.~\ref{fig:Fig_1}). 
Daily precipitation data at the four sites were obtained from EDEN, which relies on the Next Generation Weather Radar~(NEXRAD; \cite{huebner2003development}) network. Precipitation was only available for the period 2002--2021 (20 years). 
However, the data span can be considered sufficient to infer the long-term statistics of precipitation at the study sites.
EDEN also provided potential evapotranspiration ($ET_p$) data at a daily temporal resolution from 1996 to 2021.
The $ET_p$ estimates are obtained via the Priestley-Taylor method~\cite{PriestleyTaylor1972} using both solar radiation retrievals from the Geostationary Operational Environmental Satellites (GOES) mission and climate ground observations from the Florida Automated Weather Network (FAWN), the State of Florida Water Management Districts, and NOAA national network~\cite{jacobs2008satellite}.
Precipitation and evapotranspiration data were used to estimate average climatic parameters at the four sites. They include the average precipitation per event ($\alpha$, mm/day), the event frequency ($\lambda$, 1/day), and the average $ET_p$ required to parametrize the water table dynamics model introduced below and described in detail in the \textit{SI Appendix}.   
\subsection*{Shallow water table dynamics under saline conditions}
\label{sec:SWTmodel}
To better elucidate the hydrologic function of vegetation in coastal wetlands, we introduce a zero-order representation of the dynamics of the water table under saline conditions. 
Building on the approach proposed in Ref.~\cite{Laio2009}, the salinity-dependent dynamics of the saturation depth, $y$, in a homogeneous soil column can be formulated as
\begin{linenomath*}
\begin{equation}\label{yDin}
\mathcal{B} (y)\frac{dy}{dt}=
Re(y) \ - \ ET(y|C) \ + \ f_l(y),
\end{equation}
\end{linenomath*}
where $\mathcal{B}(y)$ represents the specific yield defined as the ratio between the water volume released from the storage and the corresponding variation in water level. 
$Re(y)$ is the groundwater recharge rate forced by precipitation, and $ET(y|C)$ is the evapotranspiration rate for a given level of salinization.
The function $f_l(y)$ describes the lateral flow to or from the sea, considered here the main external water body. The water table position, $\tilde{y}$, is given by the sum of saturation depth and the saturated capillary fringe, assumed to be equal to the saturated matric potential $\psi_s$, namely $\tilde{y}= y + \psi_s$. 
We note that the water table in coastal wetlands lies predominantly near the soil surface due to significant water inputs from the sea or rivers.
Consequently, the water table dynamics can be described in the framework of a shallow water table (SWT) regime~\cite{Laio2009,Tamea2010}, which implies the absence of a low soil moisture zone. 
This is a realistic assumption in tidal wetlands, where periodic seawater inundations maintain elevated soil moisture even during protracted droughts~\cite{Pezeshki2001}.
Evapotranspiration, $ET$, is defined as the sum of evaporation, $E$, and plant transpiration, $T_R$~\cite{lascano1987energy,Fatichi2017}. 
Whereas water availability does not limit $ET$ in SWT conditions~\cite{Tamea2010}, the presence of soluble salts in the soil increases the energy required by plants to uptake water, thus reducing $T_R$~\cite{Munns1993,perri2018plant,Perri2019}. 
As a result, salinity affects $ET$ through $T_R$, and the evapotranspiration rate can be defined as~\cite{Runyan2010}
  \begin{linenomath*}
\begin{equation}\label{ETpart}
ET (C)= E(C) + T_R(C)\approx
(1-{\phi}) ET_p + \phi f(C) ET_p,
\end{equation}
\end{linenomath*}
where $ET_p$ is the potential evapotranspiration, $\phi$ is a $T_R/ET_p$ ratio depending on vegetation coverage, and $f(C)$ is the limiting function accounting for the reduction in transpiration due to salinity. 
In particular, transpiration takes place at a maximum rate for salinity lower than a species-specific salinity threshold $C_T$ and linearly decreases with slope $ \beta$ for $C>C_T$.
The model proposed here represents a first step toward including the effects of salinity and plant salt tolerance in the water budget of coastal wetlands. 
The stochastic water budget in Eq.~\ref{yDin} was solved assuming a stable vegetation cover and average hydroclimatic and salinity conditions, i.e., under a steady-state hypothesis. 
The path forward requires relaxing such assumptions and fully coupling the dynamics of the water table, salinity fluctuations, and vegetation dynamics (thus allowing for regime shifts) under a non-stationary climate and sea level.
A more detailed description of the model, together with the analytical solution for the steady-state probability density function of $\tilde{y}$ and the parameter estimation methods, can be found in the companion \textit{SI Appendix}.
\subsection*{Vegetation salt tolerance parameters}
\label{sec:VegParameters}
The salt tolerance parameters ($\beta$ and $C_T$) for \textit{R. mangle} and  \textit{A. germinans} were inferred from experimental data reported in the literature. 
The transpiration of \textit{R. mangle} has been reported to be maximum at a value of salinity around 150 mol/m$^3$ of NaCl~(corresponding to about 8.8 ppt; \cite{Krauss2003,Lin1992,Lin1993}) and slowly decreases for salinity above this threshold. A substantial growth reduction is observed for salinity above 500 mol/m$^3$ ($\approx$ 30 ppt). \textit{A. germinans} is slightly more tolerant than \textit{R. mangle} with an optimal salinity at circa 170 mol/m$^3$ ($\approx$ 10 ppt) and a major growth decay for salt concentrations greater than 680 mol/m$^3$~($\approx$ 40 ppt; \cite{suarez2005salinity}). 
Based on these literature data, the salt tolerance parameters have been set to $\beta$=0.022 L/g and $C_T$=8.8 g/L for \textit{R. mangle} (Sites 1), and $\beta$=0.020 L/g and $C_T$=10 g/L for \textit{A. germinans} (Sites 2-4).
Average root depth was estimated assuming that about 80\% of the total root biomass is concentrated in the top 20 cm of soil~\cite{yates1974autoecological,Pumo2010}.
Vegetation and climatic parameters are summarized in Table S1 and S2 (\textit{SI Appendix}).
\subsubsection*{Acknowledgment}{\sf\small S.P. acknowledges the support of Princeton University's Dean for Research, High Meadows Environmental Institute, Andlinger Center for Energy and the Environment, and the Office of the Provost International Fund.
A.M. was supported by the US Department of Energy Terrestrial Ecosystem Science Program (DE) under grant no. DE-SC0020116 and by the Abu Dhabi Department of Education and Knowledge (ADEK) under grant no. AARE17-250.
Gauge height and salinity observations were obtained from the US Geological Survey and are publicly available at \url{https://waterdata.usgs.gov}. 
Post-2007 data can be downloaded directly from the US Geological Survey website, while pre-2007 data are available on request from USGS. 
Sea-level observations at Station KYWF1 -- 8724580 -- Key West, FL are available from NOAA at \url{https://tidesandcurrents.noaa.gov}. 
Rainfall and potential evapotranspiration data were provided by the Everglades Depth Estimation Network (EDEN) project (\url{https://sofia.usgs.gov/eden/}). 
The Landsat 7 mosaic and land use map of the Everglades National park in Fig.~\ref{fig:Fig_1},~\textit{PANEL 1} were obtained from the Florida Coastal Everglades Long-term Ecological Research (FCE-LTER) Program at \url{http://fcelter.fiu.edu/data/GIS/}. 
Shorelines and tree/plant outlines in Fig.~\ref{fig:Fig_2}\textit{A}--\textit{C} are courtesy of the Integration and Application Network, University of Maryland Center for Environmental Science (\url{ian.umces.edu/symbols/})
Imagery in Fig.~\ref{fig:Fig_3}~\textit{C}, \textit{F}, \textit{I} and \textit{L} were obtained from Google Earth (v. 7.3.3.7721). 
The Wolfram Mathematica code used to simulate the stochastic water table dynamics in coastal wetlands is available
at \url{https://github.com/sperriprinceton/Water-Table-Dynamics-in-Coastal-Ecosystems}.
}

\subsubsection*{Data Availability}{\sf\small Gage height and salinity observations were obtained from the US Geological Survey and are publicly available at https://waterdata.usgs.gov. Post-2007 data can be downloaded directly from the US Geological Survey website, while pre-2007 data are available on request from USGS. Sea-level observations at Station KYWF1 -- 8724580 -- Key West, FL are available from NOAA at https://tidesandcurrents.noaa.gov. Rainfall and potential evapotranspiration data were provided by the Everglades Depth Estimation Network (EDEN) project (https://sofia.usgs.gov/eden/). The Landsat 7 mosaic and land use map of the Everglades National park in Figure~\ref{fig:Fig_1}~\textbf{Panel 1} were obtained from the Florida Coastal Everglades Long-term Ecological Research (FCE-LTER) Program at http://fcelter.fiu.edu/data/GIS/. Imagery in Figure 2~\textbf{c}, \textbf{f} and \textbf{i} and \textbf{l} were obtained from Google Earth (v. 7.3.3.7721). Shorelines and tree/plant outlines in Figure~\ref{fig:Fig_3}\textbf{a}--\textbf{c} are courtesy of the Integration and Application Network, University of Maryland Center for Environmental Science (ian.umces.edu/symbols/)}

\subsubsection*{Model Availability}{\sf\small The Wolfram Mathematica code used to simulate the stochastic water table dynamics in coastal wetlands is available
at https://github.com/sperriprinceton/Water-Table-Dynamics-in-Coastal-Ecosystems.}

\subsubsection*{Competing Interests}{\sf\small The authors declare that they have no competing financial interests.}

\subsubsection*{Author Contribution}{\sf\small A.M. and S.P. conceived and designed the study. S.P. implemented the stochastic water balance model and performed the analysis; A.M. and S.P. interpreted the results and wrote the manuscript.}

\subsubsection*{Correspondence}{\sf\small Correspondence and requests for materials
should be addressed to Annalisa Molini~(email: amolini@tulane.edu) and Saverio Perri~(email: sperri@princeton.edu).}


\begin{thebibliography}{xx}

\bibitem{RamsarConventiononWetlands:2018}
{Ramsar Convention on Wetlands}, {\textit{Global wetland outlook: State of the
  World{\textquoteright}s wetlands and their services to people}}, (Institute
  for Land, Water and Society, Gland, Switzerland), Technical report (2018).

\bibitem{Bianchi2009}
TS Bianchi, MA Allison, Large-river delta-front estuaries as natural
  {\textquotedblleft}recorders{\textquotedblright} of global environmental
  change.
\newblock {\em\protect{Proc. Natl. Acad. Sci. U.S.A.}}
  \textbf{106}, 8085--8092 (2009).

\bibitem{costanza2021}
R Costanza, et~al., The global value of coastal wetlands for storm protection.
\newblock {\em\protect{Glob. Environ. Change}} \textbf{70}, 102328
  (2021).

\bibitem{Sun2020}
F Sun, RT Carson, Coastal wetlands reduce property damage during tropical
  cyclones.
\newblock {\em\protect{Proc. Natl. Acad. Sci. U.S.A.}}
  \textbf{117}, 5719--5725 (2020).

\bibitem{Feagin2009}
RA Feagin, et~al., Does vegetation prevent wave erosion of salt marsh edges?
\newblock {\em\protect{Proc. Natl. Acad. Sci. U.S.A.}}
  \textbf{106}, 10109--10113 (2009).

\bibitem{Cheng2020}
FY Cheng, KJ Van~Meter, DK Byrnes, NB Basu, {Maximizing US nitrate removal
  through wetland protection and restoration}.
\newblock {\em\protect{Nature}} \textbf{588}, 625--630 (2020).

\bibitem{Verhoeven2006}
JT Verhoeven, B Arheimer, C Yin, MM Hefting, Regional and global concerns over
  wetlands and water quality.
\newblock {\em\protect{Trends Ecol. Evol.}} \textbf{21}, 96--103
  (2006).

\bibitem{Denny1994}
P Denny, {Biodiversity and wetlands}.
\newblock {\em\protect{Wetl. Ecol. Manag.}} \textbf{3}, 55--611
  (1994).

\bibitem{rodriguez2007}
I Rodriguez-Iturbe, P D{\textquoteright}Odorico, F Laio, L Ridolfi, S Tamea,
  {Challenges in humid land ecohydrology: Interactions of water table and
  unsaturated zone with climate, soil, and vegetation}.
\newblock {\em\protect{Water Resour. Res.}} \textbf{43}, W09301
  (2007).

\bibitem{greb2006wetlands}
SF Greb, WA DiMichele, {\em Wetlands through time}.
\newblock (Geological Society of America, Boulder, CO) Vol.{} 399, (2006).

\bibitem{saenger2002mangrove}
P Saenger, {\em Mangrove ecology, silviculture and conservation}.
\newblock (Kluwer Academic Publishers, Dodrecht, The Netherlands), (2002).

\bibitem{Alongi2014}
DM Alongi, {Carbon cycling and storage in mangrove forests}.
\newblock {\em\protect{Ann. Rev. Mar. Sci.}} \textbf{6}, 195--219
  (2014).

\bibitem{McLeod2011}
E McLeod, et~al., {A blueprint for blue carbon: Toward an improved
  understanding of the role of vegetated coastal habitats in sequestering
  CO$_2$}.
\newblock {\em\protect{Front. Ecol. Environ.}} \textbf{9},
  552--560 (2011).

\bibitem{Lovelock2015}
CE Lovelock, et~al., {The vulnerability of Indo-Pacific mangrove forests to
  sea-level rise}.
\newblock 
{\em\protect{Nature}} 
\textbf{526}, 559--563 (2015).

\bibitem{Davidson2014}
NC Davidson, {How much wetland has the world lost? Long-term and recent trends
  in global wetland area}.
\newblock {\em\protect{Mar. Freshw. Res.}} \textbf{65}, 934--941
  (2014).

\bibitem{Pendleton2012}
L Pendleton, et~al., {Estimating global ``Blue Carbon" emissions from
  conversion and degradation of vegetated coastal ecosystems}.
\newblock {\em\protect{PLoS One}} \textbf{7}, e43542 (2012).

\bibitem{Jankowski2017}
KL Jankowski, TE T{\"o}rnqvist, AM Fernandes, {Vulnerability of
  Louisiana{\textquoteright}s coastal wetlands to present-day rates of relative
  sea-level rise}.
\newblock {\em\protect{Nat. Commun.}} \textbf{8}, 14792 (2017).

\bibitem{Goldberg2020}
L Goldberg, D Lagomasino, N Thomas, T Fatoyinbo, {Global declines in
  human-driven mangrove loss}.
\newblock {\em\protect{Glob. Chang. Biol.}} \textbf{26},
  5844--5855 (2020).

\bibitem{Dangendorf2019}
S Dangendorf, et~al., Persistent acceleration in global sea-level rise since
  the 1960s.
\newblock {\em\protect{Nat. Clim. Change}} \textbf{9}, 705--710
  (2019).

\bibitem{Craft2009}
C Craft, et~al., {Forecasting the effects of accelerated sea-level rise on
  tidal marsh ecosystem services}.
\newblock {\em\protect{Front. Ecol. Environ.}} \textbf{7}, 73--78
  (2009).

\bibitem{Holmquist2021}
JR Holmquist, LN Brown, GM MacDonald, {Localized scenarios and latitudinal
  patterns of vertical and lateral resilience of tidal marshes to sea-level
  rise in the contiguous United States}.
\newblock \textit{Earth's Future} \textbf{9}, e2020EF001804
  (2021).

\bibitem{Murray2022}
NJ Murray, et~al., {High-resolution mapping of losses and gains of Earth's
  tidal wetlands}.
\newblock {\em\protect{Science}} \textbf{376}, 744--749 (2022).

\bibitem{Nicholls2004}
RJ Nicholls, {Coastal flooding and wetland loss in the 21$^{st}$ century:
  Changes under the SRES climate and socio-economic scenarios}.
\newblock {\em\protect{Glob. Environ. Change}} \textbf{14}, 69--86
  (2004) Climate Change.

\bibitem{Kirwan2016a}
ML Kirwan, S Temmerman, EE Skeehan, GR Guntenspergen, S Fagherazzi,
  {Overestimation of marsh vulnerability to sea level rise}.
\newblock {\em\protect{Nat. Clim. Chang.}} \textbf{6}, 253--260
  (2016).

\bibitem{schuerch2018future}
M Schuerch, et~al., {Future response of global coastal wetlands to sea-level
  rise}.
\newblock {\em\protect{Nature}} \textbf{561}, 231--238 (2018).

\bibitem{Wiberg2019}
PL Wiberg, S Fagherazzi, ML Kirwan, {Improving predictions of salt marsh
  evolution through better integration of data and models}.
\newblock {\em\protect{Ann. Rev. Mar. Sci.}} \textbf{12} (2019).

\bibitem{Thorne2018}
K Thorne, et~al., {U.S. Pacific coastal wetland resilience and vulnerability to
  sea-level rise}.
\newblock {\em\protect{Sci. Adv.}} \textbf{4}, eaao3270 (2018).

\bibitem{Kirwan2012}
ML Kirwan, SM Mudd, {Response of salt-marsh carbon accumulation to climate
  change}.
\newblock {\em\protect{Nature}} \textbf{489}, 550--553 (2012).

\bibitem{Rogers2019}
K Rogers, et~al., {Wetland carbon storage controlled by millennial-scale
  variation in relative sea-level rise}.
\newblock {\em\protect{Nature}} \textbf{567}, 91 (2019).

\bibitem{Kirwan2016b}
ML Kirwan, DC Walters, WG Reay, JA Carr, {Sea level driven marsh expansion in a
  coupled model of marsh erosion and migration}.
\newblock {\em\protect{Geophys. Res. Lett.}} \textbf{43},
  4366--4373 (2016).

\bibitem{Kirwan2013}
ML Kirwan, JP Megonigal, {Tidal wetland stability in the face of human impacts
  and sea-level rise}.
\newblock {\em\protect{Nature}} \textbf{504}, 53--60 (2013).

\bibitem{Kirwan2010}
ML Kirwan, et~al., {Limits on the adaptability of coastal marshes to rising sea
  level}.
\newblock {\em\protect{Geophys. Res. Lett.}} \textbf{37}, L23401
  (2010).

\bibitem{morris2002}
JT Morris, PV Sundareshwar, CT Nietch, B Kjerfve, DR Cahoon, {Responses of
  coastal wetlands to rising sea level}.
\newblock {\em\protect{Ecology}} \textbf{83}, 2869--2877 (2002).

\bibitem{Mudd2009}
SM Mudd, SM Howell, JT Morris, {Impact of dynamic feedbacks between
  sedimentation, sea-level rise, and biomass production on near-surface marsh
  stratigraphy and carbon accumulation}.
\newblock {\em\protect{Estuar. Coast. Shelf Sci.}} \textbf{82},
  377--389 (2009).

\bibitem{Nordio2022}
G Nordio, S Fagherazzi, {Salinity increases with water table elevation at the
  boundary between salt marsh and forest}.
\newblock {\em\protect{J. Hydrol.}} \textbf{608}, 127576 (2022).

\bibitem{bui2013soil}
EN Bui, {Soil salinity: A neglected factor in plant ecology and biogeography}.
\newblock {\em\protect{J. Arid Environ.}} \textbf{92}, 14--25
  (2013).

\bibitem{perri2017salinity}
S Perri, F Viola, LV Noto, A Molini, Salinity and periodic inundation controls
  on the soil-plant-atmosphere continuum of gray mangroves.
\newblock {\em\protect{Hydrol. Process.}} \textbf{31}, 1271--1282
  (2017).

\bibitem{perri2018plant}
S Perri, D Entekhabi, A Molini, {Plant osmoregulation as an emergent
  water-saving adaptation}.
\newblock {\em\protect{Water Resour. Res.}} \textbf{54},
  2781--2798 (2018).

\bibitem{Craft2007}
C Craft, {Freshwater input structures soil properties, vertical accretion, and
  nutrient accumulation of Georgia and US tidal marshes}.
\newblock {\em\protect{Limnol. Oceanogr.}} \textbf{52}, 1220--1230
  (2007).

\bibitem{Neubauer2008}
SC Neubauer, {Contributions of mineral and organic components to tidal
  freshwater marsh accretion}.
\newblock {\em\protect{Estuar. Coast. Shelf Sci.}} \textbf{78},
  78--88 (2008).

\bibitem{Morris2016}
JT Morris, et~al., {Contributions of organic and inorganic matter to sediment
  volume and accretion in tidal wetlands at steady state}.
\newblock {\em{Earth's Future}} \textbf{4}, 110--121
  (2016).

\bibitem{Nyman2006}
JA Nyman, RJ Walters, RD Delaune, WH Patrick, Marsh vertical accretion via
  vegetative growth.
\newblock {\em\protect{Estuar. Coast. Shelf Sci.}} \textbf{69},
  370--380 (2006).

\bibitem{Turner2002}
RE Turner, EM Swenson, CS Milan, {Organic and inorganic contributions to
  vertical accretion in salt marsh sediments} in {\em Concepts and
  controversies in tidal marsh ecology}.
\newblock (Springer, Dordrecht, The Netherlands), pp. 583--595 (2002).

\bibitem{Herbert2015}
ER Herbert, et~al., {A global perspective on wetland salinization: Ecological
  consequences of a growing threat to freshwater wetlands}.
\newblock {\em\protect{Ecosphere}} \textbf{6}, 1--43 (2015).

\bibitem{Bathmann2021}
J Bathmann, et~al., Modelling mangrove forest structure and species composition
  over tidal inundation gradients: The feedback between plant water use and
  porewater salinity in an arid mangrove ecosystem.
\newblock {\em\protect{Agric. For. Meteorol.}} \textbf{308},
  108547 (2021).

\bibitem{Perri2020pnas}
S Perri, et~al., {River basin salinization as a form of aridity}.
\newblock {\em\protect{Proc. Natl. Acad. Sci. U.S.A.}}
  \textbf{117}, 17635--17642 (2020).

\bibitem{Perri2019}
S Perri, GG Katul, A Molini, {Xylem‐ phloem hydraulic coupling explains
  multiple osmoregulatory responses to salt‐stress}.
\newblock {\em\protect{New Phytol.}} \textbf{224}, 644--662
  (2019).

\bibitem{Lin1993}
GH Lin, L Sternberg, {Effects of salinity fluctuation on photosynthetic gas
  exchange and plant growth of the red mangrove (\textit{Rhizophora mangle
  L.})}.
\newblock {\em\protect{J. Exp. Bot.}} \textbf{44}, 9--16 (1993).

\bibitem{Suarez2006}
N Suarez, E Medina, {Influence of salinity on {Na$^+$} and {K$^+$}
  accumulation, and gas exchange in \textit{Avicennia germinans}}.
\newblock {\em\protect{Photosynthetica}} \textbf{44}, 268--274
  (2006).

\bibitem{munns2008mechanisms}
R Munns, M Tester, {Mechanisms of salinity tolerance}.
\newblock {\em\protect{Annu. Rev. Plant Biol.}} \textbf{59},
  651--681 (2008).

\bibitem{Dacey1984}
JWH Dacey, BL Howes, {Water uptake by roots controls water table movement and
  sediment oxidation in short \textit{Spartina} marsh}.
\newblock {\em\protect{Science}} \textbf{224}, 487--489 (1984).

\bibitem{Dube1995}
S Dub{\'e}, AP Plamondon, RL Rothwell, {Watering up after clear-cutting on
  forested wetlands of the St. Lawrence lowland}.
\newblock {\em\protect{Water Resour. Res.}} \textbf{31},
  1741--1750 (1995).

\bibitem{Roy2000}
V Roy, JC Ruel, AP Plamondon, Establishment, growth and survival of natural
  regeneration after clearcutting and drainage on forested wetlands.
\newblock {\em\protect{For. Ecol. Manag.}} \textbf{129}, 253--267
  (2000).

\bibitem{Ewe2006}
SML Ewe, et~al., {Spatial and temporal patterns of aboveground net primary
  productivity (ANPP) along two freshwater-estuarine transects in the Florida
  Coastal Everglades}.
\newblock {\em\protect{Hydrobiologia}} \textbf{569}, 459--474
  (2006).

\bibitem{Poret2007}
N Poret, RR Twilley, VH Rivera-Monroy, C Coronado-Molina, {Belowground
  decomposition of mangrove roots in Florida coastal everglades}.
\newblock {\em\protect{Estuaries Coasts}} \textbf{30}, 491--496
  (2007).

\bibitem{Wang2011}
W Wang, et~al., {Mangroves: Obligate or facultative halophytes? A review}.
\newblock {\em\protect{Trees}} \textbf{25}, 953--963 (2011).

\bibitem{Davis2001}
S Davis, D Childers, J Day, D Rudnick, F Sklar, {Nutrient dynamics in vegetated
  and unvegetated areas of a southern Everglades mangrove creek}.
\newblock {\em\protect{Estuar. Coast. Shelf Sci.}} \textbf{52},
  753--768 (2001).

\bibitem{Maliva2021}
R Maliva, {\em Sea level rise and groundwater}.
\newblock (Springer International Publishing), pp. 113--153 (2021).

\bibitem{Befus2020}
KM Befus, PL Barnard, DJ Hoover, JA Finzi~Hart, CI Voss, {Increasing threat of
  coastal groundwater hazards from sea-level rise in California}.
\newblock {\em\protect{Nat. Clim. Change}} \textbf{10}, 946--952
  (2020).

\bibitem{Perri:2018ur}
S Perri, S Suweis, D Entekhabi, A Molini, {Vegetation controls on dryland
  salinity}.
\newblock {\em\protect{Geophys. Res. Lett.}} \textbf{45}, 669--682
  (2018).

\bibitem{perri2022natgeo}
S Perri, A Molini, LO Hedin, A Porporato, Contrasting effects of aridity and
  seasonality on global salinization.
\newblock {\em\protect{Nat. Geosci.}} \textbf{15}, 375–--381
  (2022).

\bibitem{Scheffer2001}
M Scheffer, S Carpenter, JA Foley, C Folke, B Walker, Catastrophic shifts in
  ecosystems.
\newblock {\em\protect{Nature}} \textbf{413}, 591--596 (2001).

\bibitem{Pumo2010}
D Pumo, S Tamea, LV Noto, F Miralles-Wilhem, I Rodriguez-Iturbe, {Modeling
  belowground water table fluctuations in the Everglades}.
\newblock {\em\protect{Water Resour. Res.}} \textbf{46}, W11557
  (2010).

\bibitem{Laio2009}
F Laio, S Tamea, L Ridolfi, P D'Odorico, I Rodriguez-Iturbe, {Ecohydrology of
  groundwater-dependent ecosystems: 1. Stochastic water table dynamics}.
\newblock {\em\protect{Water Resour. Res.}} \textbf{45}, W05419
  (2009).

\bibitem{Tamea2009}
S Tamea, F Laio, L Ridolfi, P D'Odorico, I Rodriguez-Iturbe, {Ecohydrology of
  groundwater-dependent ecosystems: 2. Stochastic soil moisture dynamics}.
\newblock {\em\protect{Water Resour. Res.}} \textbf{45}, W05420
  (2009).

\bibitem{Tamea2010}
S Tamea, R Muneepeerakul, F Laio, L Ridolfi, I Rodriguez-Iturbe, {Stochastic
  description of water table fluctuations in wetlands}.
\newblock {\em\protect{Geophys. Res. Lett.}} \textbf{37}, L06403
  (2010).

\bibitem{maas1999crop}
E Maas, S Grattan, {\em Crop yields as affected by salinity}.
\newblock (American Society of Agronomy, Madison, WI), pp. 55--108 (1999).

\bibitem{richardson2008everglades}
C Richardson, {\em {The Everglades experiments: Lessons for ecosystem
  restoration}}.
\newblock (Springer, New York City, NY), (2008).

\bibitem{passioura1992mangroves}
JB Passioura, MC Ball, JH Knight, {Mangroves may salinize the soil and in so
  doing limit their transpiration rate}.
\newblock {\em\protect{Funct. Ecol.}}, 476--481 (1992).

\bibitem{Wendelberger2017}
KS Wendelberger, JH Richards, {Halophytes can salinize soil when competing with
  glycophytes, intensifying effects of sea level rise in coastal communities}.
\newblock {\em\protect{Oecologia}} \textbf{184}, 729--737 (2017).

\bibitem{Lagomasino2021}
D Lagomasino, et~al., {Storm surge and ponding explain mangrove dieback in
  southwest Florida following Hurricane Irma}.
\newblock {\em\protect{Nat. Commun.}} \textbf{12}, 4003 (2021).

\bibitem{jones2019}
MC Jones, et~al., {Rapid inundation of southern Florida coastline despite low
  relative sea-level rise rates during the late-Holocene}.
\newblock {\em\protect{Nat Commun.}} \textbf{10}, 3231 (2019).

\bibitem{Smith2009}
TJ Smith, et~al., {Cumulative impacts of hurricanes on Florida mangrove
  ecosystems: Sediment deposition, storm surges and vegetation}.
\newblock {\em\protect{Wetlands}} \textbf{29}, 24 (2009).

\bibitem{langevin2005simulation}
C Langevin, E Swain, M Wolfert, Simulation of integrated
  surface-water/ground-water flow and salinity for a coastal wetland and
  adjacent estuary.
\newblock {\em\protect{J. Hydrol.}} \textbf{314}, 212--234 (2005).

\bibitem{satbenau2012salinity}
E Satbenau, K Kotun, {Salinity and hydrology of Florida Bay: Status and trends
  1990-2009}, (National Park Service, South Florida Natural Resources Center,
  Homestead, FL), Technical report (2012).

\bibitem{Marshall2014}
FE Marshall, GL Wingard, PA Pitts, {Estimates of natural salinity and hydrology
  in a subtropical estuarine ecosystem: Implications for Greater Everglades
  restoration}.
\newblock {\em\protect{Estuaries Coast.}} \textbf{37}, 1449--1466
  (2014).

\bibitem{national2019progress}
{National Academies of Sciences, Engineering}, {Medicine}, {\em Progress toward
  restoring the Everglades: The seventh biennial review - 2018}.
\newblock (The National Academies Press, Washington, DC), (2019).

\bibitem{Peel2007}
MC Peel, BL Finlayson, TA McMahon, {Updated world map of the K{\"o}ppen-Geiger
  climate classification}.
\newblock {\em\protect{Hydrol. Earth Syst. Sci.}} \textbf{11},
  1633--1644 (2007).

\bibitem{Duever1994}
MJ Duever, JF Meeder, LC Meeder, JM McCollom, {The climate of south Florida and
  its role in shaping the Everglades ecosystem} in {\em Everglades: The
  ecosystem and its restoration}.
\newblock (CRC Press, Boca Raton, FL), pp. 225--248 (1994).

\bibitem{Ewe2007}
SM Ewe, LDS Sternberg, DL Childers, {Seasonal plant water uptake patterns in
  the saline southeast Everglades ecotone}.
\newblock {\em\protect{Oecologia}} \textbf{152}, 607--616 (2007).

\bibitem{Wahl2014}
T Wahl, FM Calafat, ME Luther, {Rapid changes in the seasonal sea level cycle
  along the US Gulf coast from the late 20$^{th}$ century}.
\newblock {\em\protect{Geophys. Res. Lett.}} \textbf{41}, 491--498
  (2014).

\bibitem{Koch1997}
MS Koch, SC Snedaker, {Factors influencing\textit{ Rhizophora mangle} L.
  seedling development in Everglades carbonate soils}.
\newblock {\em\protect{Aquat. Bot.}} \textbf{59}, 87--98 (1997).

\bibitem{Barr2009}
JG Barr, JD Fuentes, V Engel, JC Zieman, {Physiological responses of red
  mangroves to the climate in the Florida Everglades}.
\newblock {\em\protect{J. Geophys. Res. Biogeosci.}} \textbf{114},
  G02008 (2009).

\bibitem{Loveless1959}
CM Loveless, {A study of the vegetation in the Florida Everglades}.
\newblock {\em\protect{Ecology}} \textbf{40}, 1--9 (1959).

\bibitem{Ross2000}
MS Ross, JF Meeder, JP Sah, PL Ruiz, GJ Telesnicki, {The Southeast Saline
  Everglades revisited: 50 years of coastal vegetation change}.
\newblock {\em\protect{J. Veg. Sci.}} \textbf{11}, 101--112
  (2000).

\bibitem{Egler1952}
FE Egler, {Southeast saline Everglades vegetation, Florida and its management}.
\newblock {\em\protect{Veg. Acta Geobot.}} \textbf{3}, 213--265
  (1952).

\bibitem{huebner2003development}
RS Huebner, C Pathak, BC Hoblit, {\em Development and use of a NEXRAD database
  for water management in South Florida}.
\newblock (Proc., ASCE EWRI World Water and Environmental Resources Congress,
  Reston, VA) Vol.{} 118, pp. 1--10 (2003).

\bibitem{PriestleyTaylor1972}
CHB Priestley, RJ Taylor, {On the assessment of surface heat flux and
  evaporation using large-scale parameters}.
\newblock {\em\protect{Mon. Weather Rev.}} \textbf{100}, 81--92
  (1972).

\bibitem{jacobs2008satellite}
J Jacobs, J Mecikalski, S Paech, {Satellite-based solar radiation, net
  radiation, and potential and reference evapotranspiration estimates over
  Florida}.
\newblock {\em\protect{US Geological Survey, Orlando, FL}} (2008).

\bibitem{Pezeshki2001}
SR Pezeshki, {Wetland plant responses to soil flooding}.
\newblock {\em\protect{Environ. Exp. Bot.}} \textbf{46}, 299--312
  (2001).

\bibitem{lascano1987energy}
R Lascano, C Van~Bavel, J Hatfield, D Upchurch, Energy and water balance of a
  sparse crop: Simulated and measured soil and crop evaporation.
\newblock {\em\protect{Soil Sci. Soc. Am. J.}} \textbf{51},
  1113--1121 (1987).

\bibitem{Fatichi2017}
S Fatichi, C Pappas, {Constrained variability of modeled T:ET ratio across
  biomes}.
\newblock {\em\protect{Geophys. Res. Lett.}} \textbf{44},
  6795--6803 (2017).

\bibitem{Munns1993}
R Munns, {Physiological processes limiting plant-growth in saline soils -- Some
  dogmas and hypotheses}.
\newblock {\em\protect{Plant Cell Environ.}} \textbf{16}, 15--24
  (1993).

\bibitem{Runyan2010}
CW Runyan, P D'Odorico, {Ecohydrological feedbacks between salt accumulation
  and vegetation dynamics: Role of vegetation-groundwater interactions}.
\newblock {\em\protect{Water Resour. Res.}} \textbf{46}, W11561
  (2010).

\bibitem{Krauss2003}
KW Krauss, JA Allen, {Influences of salinity and shade on seedling
  photosynthesis and growth of two mangrove species, \textit{Rhizophora mangle}
  and \textit{Bruguiera sexangula}, introduced to Hawaii}.
\newblock {\em\protect{Aquat. Bot.}} \textbf{77}, 311--324 (2003).

\bibitem{Lin1992}
GH Lin, L Sternberg, {Effect of growth form, salinity, nutrient and sulfide on
  photosynthesis, carbon isotope discrimination and growth of red mangrove
  (\textit{Rhizophora mangle} L.)}.
\newblock {\em\protect{Funct. Plant Biol.}} \textbf{19}, 509--517
  (1992).

\bibitem{suarez2005salinity}
N Su{\'a}rez, E Medina, Salinity effect on plant growth and leaf demography of
  the mangrove, \textit{Avicennia germinans} l.
\newblock {\em\protect{Trees}} \textbf{19}, 722 (2005).

\bibitem{yates1974autoecological}
SA Yates, An Autoecological Study of Sawgrass: Cladium Jamaicense, in South Florida, Ph.D. thesis, University of Miami, Miami, FL (1974).

\end{thebibliography}
\end{document}